# Physics Informed Multi-task Joint Generative Learning for Arterial Vehicle Trajectory Reconstruction Considering Lane Changing Behavior

Mengyun Xu[a,b], Jie Fang[b]*, Eui-Jin Kim[c]*, Tony Z. Qiu[d,e], Prateek Bansal[a]

[a]Department of Civil and Environmental Engineering, National University of Singapore, Singapore 117576, Singapore.
[b]Department of Civil Engineering, Fuzhou University, Fuzhou ,350108, China
[c]Department of Transportation System Engineering, Ajou University, Suwon 16499, Republic of Korea
[d] Department of Civil and Environmental Engineering, University of Alberta, Edmonton, AB T6G 1H9, Canada
[e] the Intelligent Transport System Research Center, Wuhan University of Technology, Wuhan, 430063 China

## Abstract

Reconstructing complete traffic flow time-space diagrams from vehicle trajectories offer a comprehensive view on traffic dynamics at arterial intersections. However, obtaining full trajectories across networks is costly, and accurately inferring lane-changing (LC) and car-following behaviors in multi-lane environments remains challenging. This study proposes a generative framework for arterial vehicle trajectory reconstruction that jointly models lane-changing and car-following behaviors through physics-informed multi-task joint learning. The framework consists of a Lane-Change Generative Adversarial Network (LC-GAN) and a Trajectory-GAN. The LC-GAN models stochastic LC behavior from historical trajectories while considering physical conditions of arterial intersections, such as signal control, geometric configuration, and interactions with surrounding vehicles. The Trajectory-GAN then incorporates LC information from the LC-GAN with initial trajectories generated from physics-based car-following models, refining them in a data-driven manner to adapt to dynamic traffic conditions. The proposed framework is designed to reconstruct complete trajectories from only a small subset of connected vehicle (CV) trajectories; for example, even a single observed trajectory per lane, by incorporating partial trajectory information into the generative process. A multi-task joint learning facilitates synergistic interaction between the LC-GAN and Trajectory-GAN, allowing each component to serves as both auxiliary supervision and a physical condition for the other. Validation using two heterogeneous real-world trajectory datasets demonstrates that the framework outperforms conventional benchmark models in reconstructing complete time-space diagrams for multi-lane arterial intersections. This research advances the integration of trajectory-based sensing from CVs with physics-informed deep learning, offering new insights for traffic management and intersection control optimization.

*Keywords:* Vehicle trajectory reconstruction, Lane-changing behavior modeling, Generative adversarial networks, Multi-task joint learning, Physics-informed deep learning

## 1. Introduction

High-fidelity vehicle trajectory reconstruction at arterial intersections is essential for comprehensive traffic state estimation, signal control optimization, and coordinated control of connected vehicles (Li et al., 2020). Reconstructed trajectories enable a transition from isolated vehicle-level control to system-wide traffic awareness, establishing the foundation for cooperative path planning. Despite advancements in trajectory-based sensing technologies, obtaining complete trajectories remain challenging. Point-based sensors such as traffic cameras can technically capture complete trajectories but requires dense and costly camera deployments with precise coverage angles, which are often impractical for large-scale arterial monitoring and financially infeasible in most countries. Alternatively, connected vehicles (CVs) act as mobile sensors, providing high-resolution data on position, velocity, and acceleration through onboard telematics. However, reconstructing complete traffic trajectories in mixed traffic conditions, where CV penetration is low and human-driven vehicles (HVs) dominate, remain a critical challenge. To address this, this study develops a framework that leverages sparse CV trajectory data to infer the movement of surrounding vehicles and reconstruct the complete space-time diagram, offering a cost-effective and scalable solution for intelligent transportation systems.

Urban arterial approaches with multiple lanes exhibit complex traffic dynamics characterized by (i) stochastic vehicle arrivals induced by frequent signal phase transitions, (ii) prevalent LC behavior driven by turning demands or efficiency-seeking behavior, and (iii) significant heterogeneity in both LC and car-following dynamics due to constrained distances between intersections (Toledo et al., 2007).

Existing research on trajectory reconstruction in arterial environments has primarily focused on estimating vehicle arrival distributions and modeling single-lane car-following behavior, with limited capability in capturing LC dynamics. LC behavior, however, serves as a crucial spatiotemporal anchor that reflects a vehicle's lateral transitions across the adjacent lanes, typically for improving driving conditions or making a turn. It is thus essential to achieving physically consistent spatiotemporal trajectory reconstruction. In practice, LC behavior exhibits dual complexity: at the individual-vehicle level, it requires the joint consideration of lateral maneuvers and longitudinal car-following dynamics; at the multi-vehicle interaction level, it involves complex game-theoretic behaviors. Neglecting LC dynamics hinders both the physical plausibility and practical applicability of multi-lanes trajectory reconstruction. This study aims to reconstruct complete vehicle trajectories from sparse CV trajectory data by explicitly modeling LC behavior along with car following behavior and capturing comprehensive spatiotemporal movement patterns.

Current LC behavior models predominantly applied on freeway scenarios, which exhibit relatively straightforward traffic patterns, and are thus inadequate for arterial applications due to three fundamental challenges. First, under signalized constraints, the impact of traffic signals on LC behavior remains poorly quantified, particularly for mandatory lane changes (MLCs) that must occur within limited green windows to satisfy turning requirements. Second, the geometric complexity of arterial networks, characterized by restricted spacing between consecutive intersections, severely limits available LC distances, rendering conventional rule-based or fixed assumptions models ineffective in capturing heterogeneous driving behaviors. Third, the coexistence of MLCs and discretionary lane changes (DLCs) across diverse spatiotemporal contexts introduces strong condition-dependent variations, which fixed-parameter models fail to represent effectively.

The key challenge in LC behavior modeling is the probabilistic inference of spatiotemporal coordinates that determine when and where a vehicle initiates a LC. This is a stochastic process conditioned on interactions with surrounding vehicles, signal control, and geometric configuration. Conditional Generative Adversarial Networks (cGANs) offer a promising paradigm for this task by leveraging adversarial learning to capture complex conditional dependencies embedded in trajectory data. Building on this concept, this study develops tailored cGAN architecture to improve the accuracy and physical consistency of LC behavior modeling.

The LC position (LCP) determines the breakpoints in a vehicle's trajectory as it transitions across adjacent lanes, while the kinematic constraints governing trajectory inference determine the feasible spatiotemporal domains for LC behavior. This phenomenon exhibits a strong bidirectional coupling between LC behavior modeling and trajectory reconstruction. Independent modeling of these behaviors induces error propagation and behavioral distortions due to inadequate representation of their spatiotemporal interdependencies. To address this coupling challenge, we propose a Multi-task Joint Generative Learning-based Trajectory Reconstruction Framework (MGL-TRF) with explicit incorporation of LC dynamics.

The key contributions of this study are as follows:

- The proposed MGL-TRF establishes a physical-informed deep generative modeling framework to reconstruct vehicle trajectories in complex urban arterial networks. Leveraging a multi-task learning technique, this study proposes the integrated framework to jointly model the car-following and LC behaviors for arterial vehicle trajectory reconstruction. Each component serves simultaneously as auxiliary supervision and physical condition for the other, thereby enforcing behavioral consistency and significantly reducing reconstruction uncertainty relative to single-task baselines.

- The framework embeds key physical conditions, such as interactions with surrounding vehicles, signal control, and road geometrics, into data-driven LC behavior modeling. These conditions effectively regularize a deep generative model, enabling it to accommodate both MLC and DLC across diverse intersection configurations compared to conventional models.

- We validate and benchmark the proposed MGL-TRF using two real-world vehicle trajectory datasets, confirming its effectiveness and superiority in reconstructing complete time–space diagrams for multi-lane arterial intersections.

## *2.* Literature review

### 2.1. LC Estimation Model

LC estimation modeling represents a significant branch of transportation research, with numerous approaches developed over the past decades. These models can be broadly categorized into two groups: physics-based models, including rule-based and utility-based models (Kesting et al., 2007; Choudhury and Ben-Akiva, 2013), and data-driven models employing machine learning techniques such as neural networks (Gao et al., 2018) and support vector machines (SVM) (Kumar et al., 2013).

Rule-based LC models typically address lane selection and gap acceptance through deterministic decision rules (Choudhury and Ben-Akiva, 2013; Gipps et al., 1986). Subsequent refinements have incorporated utility-based structure considering the trade-offs between safety and efficiency (Laval and Leclercq, 2008). For instance, Toledo et al. (2003) formulated a unified utility function applicable to both MLC and DLC, while Kesting et al. (2007) introduced the Minimizing Overall Braking Induced by Lane changes (MOBIL) model, which integrates LC benefits and risks within an acceleration-based utility function derived from car-following theory. Sun and Elefteriadou (2012) further extended the utility-based approach by incorporating driver heterogeneity and various contextual factors into the utility specification.

Most of these studies have focused on freeway or highway environments, with limited applications to urban arterials, where signal control and geometrics affect complicate LC decision processes and the estimation of LCP distributions. Arterial traffic flow is frequently disrupted by periodic signal phases, resulting in complex stop-and-go dynamics. Sun and Elefteriadou (2014) demonstrated that incorporating driver characteristics into a utility-based LC model improves behavioral realism in urban contexts. Nevertheless, model generalizability across different intersections remains challenging due to the diversity of local traffic and signal conditions. Furthermore, both rule-based and utility-based LC models often overlook complex arterial-specific factors such as interactions among surrounding vehicles, driver heterogeneity, and the impacts of signal timing and network geometry.

In contrast, data-driven LC models aim to extract behavioral patterns directly from empirical data rather than relying on pre-defined physical assumptions. Dou et al. (2016) developed an MLC prediction model for highway lane drops using neural networks and SVM using features such as speed differences, vehicle gaps, and positional information. Kumar et al. (2013) proposed an SVM-based LC behavior predictor using Bayesian filtering. Recent advances in deep learning have further enhanced the predictability of such models. Xie et al. (2019) applied deep belief networks (DBN) to model LC decision-making processes. While data-driven models excel at capturing complex, non-linear behavioral patterns that are difficult to represent through physics-based rules, they typically require large-scale datasets and may struggle with limited generalizability across heterogeneous traffic contexts. To address these limitations, recent research has explored the integration of physical conditions into data-driven architectures. Michail et al. (2023) developed an adaptive physics-informed trajectory reconstruction framework that incorporates driver behavior and vehicle dynamics, achieving a significant reduction in speed estimation error while maintaining consistent performance across different data acquisition systems. Chen et al. (2024) developed an interactive multi-model framework that integrates a Gaussian process-based motion modeling with a physics-based trajectory prediction using an Extended Kalman Filter, significantly improving the accuracy of lane change trajectory prediction for autonomous vehicles. The Physics-Informed Generative Adversarial Network (PIGAN) introduced by Zhou et al. (2021) embeds governing equations that describe system reliability evolution as regularization terms within the loss function, thereby improving the physical consistency and robustness of reliability predictions. These related studies have motivated hybrid approaches that integrate physical conditions within data-driven architectures to improve behavioral fidelity and robustness in LC representation.

### 2.2. Trajectory Reconstruction Model

Existing research has focused on reconstructing vehicle trajectories at signalized arterial intersections to generate a complete spatiotemporal traffic flow diagram, accounting for both vehicle-to-vehicle interactions and intersection-induced movement patterns (Uhlemann, 2016; Yang et al., 2018; Sun and Ban, 2013). In recent years, theory-driven approaches have also been applied in the CAVs environment. Chen et al. (2022b) applied classical shockwave theory to identify queue positions and utilized the Intelligent Driver Model (IDM) and its variants to infer sub-trajectories across different spatiotemporal regions. Mehran et al. (2012) employed macroscopic kinematic wave theory to reconstruct vehicle trajectories along arterial corridors, capturing entry and exit patterns at intersections. However, such approaches often overlook the stochastic characteristic of traffic flow, resulting in limited reconstruction accuracy. To address this, Chen et al. (2021) introduced a hybrid method for reconstructing vehicle trajectories at signalized intersections that explicitly account for stochastic queue dynamics. This approach integrates physical Kalman filtering to capture and reproduce the probabilistic characteristics of queue boundary evolution, thereby enhancing the accuracy of reconstructed trajectories.

Although these theory-driven methods account for multi-vehicle dynamics and sequential intersection effects through theory-driven constructs, such as kinematic wave propagation and physics-based car-following models, they often rely on a single set of parameters calibrated for average driving behavior. This simplification limits their ability to represent the heterogeneity and stochasticity observed in real-world traffic (Makridis et al., 2023). To overcome this limitation, recent studies have explored data-driven approaches. For instance, Wang et al. (2022) suggested leveraging data-driven approaches to address the vehicle trajectory reconstruction problem with minimal assumptions is promising with the tremendous data available. Wang et al. (2024) proposed IDM-Follower, a data-driven approach that employs a dual-encoder with attention-decoder recurrent auto encoder architecture constrained by the IDM to predict car-following trajectories. Xu et al. (2025) proposed a trajectory reconstruction framework that integrates a GAN for arrival distribution estimation with another GAN for trajectory generation, achieving accurate estimation of complete spatiotemporal trajectories.

Nevertheless, existing trajectory reconstruction models for arterial networks usually fail to adequately capture LC dynamics. This limitation persists even in simpler freeway contexts, where LC prediction and trajectory reconstruction are commonly treated as sequential rather than integrated processes. For instance, Xue et al. (2022) developed an integrated LC model that combines XGBoost for LC decision prediction with an LSTM for trajectory generation, while Xie et al. (2019) applied Deep Belief Networks (DBN) and LSTM to model the decision and execution phases of LC behavior. However, decoupling LC behavior modeling from continuous trajectory reconstruction undermine temporal consistency and spatial accuracy, thereby limiting the reliability of complete multi-lane trajectory reconstruction in urban arterial environments. Meng et al. (2023) developed a multi-task learning framework based on LSTM architecture to simultaneously predict lateral trajectory, longitudinal trajectory, and lane-changing behavior. However, the model was primarily trained and validated using extensive simulation data under idealized conditions, which limiting its generalizability to real-world environment with complex dynamic interactions and limited data. In such scenarios, purely data-driven trajectory prediction approaches without physical conditions may struggle to effectively differentiate between lane-changing behavior modeling and trajectory reconstruction tasks.

## 2.3. Application of GANs for Trajectory Data

Recent advances in deep generative modeling, particularly GANs, have demonstrated remarkable capabilities in capturing the probability distributions of stochastic processes (Odena et al., 2017). cGANs extend this capacity by integrating conditional inputs that reflect physical constraints.

In trajectory reconstruction, Wang et al. (2021) introduced a two-stage GAN framework that effectively combines mobility pattern with geographical features to generate continuous GPS trajectories. Similarly, Dendorfer et al. (2021) proposed a multi-generator structure to mitigate out-of-distribution sampling issues in pedestrian trajectory forecasting. These studies underscore potential of GAN to learn complex trajectory data distributions. In the context of signalized arterials, Xu et al. (2025) developed an Arrival-GAN for estimating vehicle arrival distributions and a physics-informed Trajectory-GAN for reconstructing complete trajectories. The demonstrated success of GANs in modeling trajectory data motivates their adaptation for LC estimation and trajectory reconstruction at arterial intersections. However, a critical challenge that remains is the effective integration of LC prediction and trajectory reconstruction into a unified framework.

## 2.4. Summary and Research Gap

While substantial progress has been made in applying physics-informed deep learning models for car-following and lane-changing prediction for trajectory reconstruction, a unified framework that jointly addresses both tasks in urban arterials remains lacking. Existing studies that consider both tasks are largely limited to freeway scenarios or simulated datasets and fail to fully exploit their interdependence.

To bridge this gap, we propose a Multi-tasks joint Generative Learning based Trajectory Reconstruction Framework (MGL-TRF) that unifies physical conditions with data-driven learning and integrate LC estimation with multi-lane trajectory reconstruction at signalized arterials. The framework decomposes complete spatiotemporal traffic flow reconstruction into two coupled tasks: LC behavior modeling and trajectory reconstruction. First, a physics-informed LC-GAN captures stochastic LC dynamics influenced by signal controls, geometric configuration, and surrounding vehicle interactions, factors often underrepresented by conventional rule- or utility-based models. Second, a jointly trained framework incorporates LC outputs into a Trajectory-GAN to ensure behavioral and kinematic consistency across lanes. This co-training approach preserves the interpretability of physical models while leveraging the flexibility of deep generative models, thereby improving the accuracy and practical applicability of trajectory reconstruction in arterial networks under low CV penetration conditions.

## *3*. Methodology

## 3.1. Problem Definition

This study aims to reconstruct complete vehicle trajectories at arterial intersections with LC behavior using partially available high-resolution CV trajectory data through a physics-informed multi-task joint generative learning. The reconstruction process assumes that at least one CV-equipped vehicle per lane remain its lane, providing a physically plausible reference for inferring complete trajectories of all following vehicles and allowing effective reconstruction even under extremely low CV penetration conditions.

To facilitate the inference of LC vehicles (LCVs), it is further assumed that conventional fixed detectors are deployed at both the segment entrance and near the maximum queue length. These detectors record vehicle arrival/departure timestamps and vehicle counts, enabling the derivation of critical input features for the proposed framework. The problem is formally defined in Equation 1.

$$\begin{pmatrix} ..., \boldsymbol{hTraj}_1^{i-1}, \boldsymbol{cTraj}_1^i, \boldsymbol{hTraj}_1^{i+1}, ..., \boldsymbol{cTraj}_1^k, ..., \boldsymbol{hTraj}_1^n \\ \vdots \\ ..., \boldsymbol{hTraj}_l^{i-1}, \boldsymbol{cTraj}_l^i, \boldsymbol{hTraj}_l^{i+1}, ..., \boldsymbol{cTraj}_l^k, ..., \boldsymbol{hTraj}_l^n \end{pmatrix} = F \begin{pmatrix} \boldsymbol{cTraj}_1^i, ..., \boldsymbol{cTraj}_1^k \\ \vdots \\ \boldsymbol{cTraj}_l^i, ..., \boldsymbol{cTraj}_l^k \end{pmatrix} \qquad (1)$$

where $\boldsymbol{cTraj}_l^i$ and $\boldsymbol{hTraj}_l^n$ are the trajectory of $i$th CV and $n$th HV in the lane $l$, respectively. For each CV trajectory, we have the information on stationing (i.e., the distance from the origin points of the roadway) and velocity.

The overall architecture of the proposed MGL-TRF is shown in Fig. 1. First, the *Lane Change Vehicles Deducing Module* take as inputs the available data, such as CV trajectory, arrival and departure time collected from detectors, and signal phase, to identify potential LC vehicles by comparing the predicted trajectory endpoints with observed vehicle positions across adjacent lanes. The identified LCVs are then fed into the *Lane Change Behavior Module* (LC-GAN), which estimates the precise LC position using a conditional GAN architecture consisting of a generator (**LC_G**) and a discriminator (**LC_D**). The LC-GAN embeds both physical car-following and LC models as kinematic conditions, generating initial trajectories with identified LC positions that reflect physically consistent vehicle behavior.

Mengyun Xu, Jie Fang, Eui-Jin Kim, Tony Z. Qiu, Prateek Bansal

Next, the *Trajectory Reconstruction Module* (Trajectory-GAN), a conditional GAN architecture consisting of a generator (**Traj_G**) and a discriminator (**Traj_D**), refines these trajectories through two distinct pathways: (1) for LC vehicles, it takes the initial trajectories with estimated LC positions as conditional inputs; and (2) for non-LC vehicles, it utilizes physics-based initial trajectories derived from car-following models as conditional inputs.

Notably, the framework processes vehicles sequentially within each lane in temporal order, initializing with leading CVs as known trajectory inputs. Each reconstructed trajectory subsequently serves as ahead vehicle trajectories (AVT) for downstream LC-GAN and Trajectory-GAN processing of following vehicles, ensuring temporal and behavioral consistency throughout the reconstruction chain.

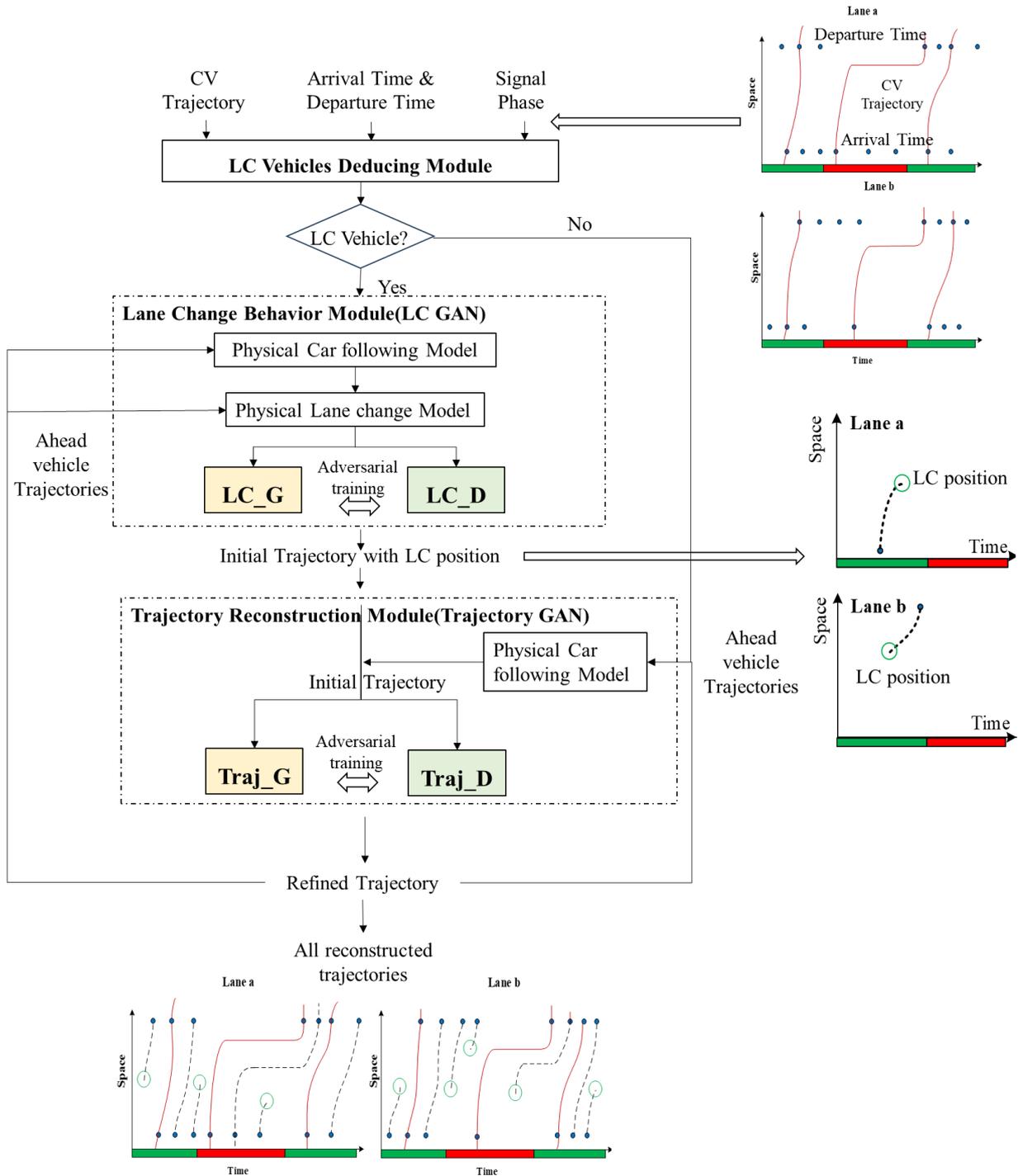

**Fig. 1.** The architecture of the proposed Multi-GANs Trajectory Reconstruction Framework (MGL-TRF)

LC vehicle identification is performed by examining the consistency between a vehicle's entry and exit lanes. Detectors positioned at the segment entrance and near the maximum queue length provide precise passing timestamps, a configuration that can be easily satisfied at most signalized urban intersections. The data processing procedure begins with the reconstruction of virtual trajectories for all lanes based on the detected vehicle counts and arrival-departure timestamps, using a physical car-following model, as detailed in Section 3.2.1. Each real departure timestamp is then matched to its temporally closest virtual trajectory endpoint to determine LC occurrence and corresponding lane pairs. The complete matching procedure, detailed in **Algorithm 1**, comprises four steps.

In the first stage (Step I), the algorithm performs intra-lane matching between downstream detection points ($D_c$) and inferred trajectory endpoints ($D_i$) to identify non-LC vehicles under pre-queue conditions. Subsequently (Step II), inter-lane matching across adjacent lanes is conducted through cross-comparison of $D_c$ and $D_i$ to detect LC behaviors occurring before queue formation. Following these two steps, the remaining unmatched detection points predominantly represent vehicles affected by queuing effects.

The algorithm then proceeds with a secondary analysis (Step III), conducting renewed intra-lane temporal matching between upstream detection points ($O_c$) and $D_c$ within each lane. Vehicles that remain unmatched after this step are provisionally classified as potential LC cases. Finally, cross-lane temporal matching (Step IV) is applied to search for chronologically proximate $O_c$-$D_c$ pairs across adjacent lanes. An iterative validation mechanism ensures physical feasibility, whenever a matching result violates physical movement constraints, the algorithm automatically reverts to Step III for recalibration. This cyclical refinement continues until all vehicle trajectories are matched in a physically plausible manner.

The algorithm therefore proceeds with secondary analysis (Step III), conducting renewed intra-lane temporal matching of $O_c$ and $D_c$ within each lane, with persistent unmatched points provisionally classified as potential LC cases. The final stage (Step IV) implements cross-lane temporal matching, searching for chronologically proximate $O_c$-$D_c$ pairs across adjacent lanes while incorporating an iterative validation mechanism - whenever a matching result violates physical movement constraints, the algorithm automatically reverts to Step III for recalibration. This cyclical refinement continues until achieving complete and physically plausible matching for all vehicles in the dataset.

Upon completion, the algorithm successfully determines critical trajectory parameters for all vehicles. For non-LCVs, the origin and destination points are precisely identified, while LCVs are additionally characterized by their original lane (O) and target lane (T) identifiers, forming O-T lane pairs. This comprehensive identification provides an essential foundation for subsequent high-fidelity trajectory reconstruction, ensuring both spatial accuracy and behavioral plausibility in the final output.

| Algorithm 1 Data processing for LCVs deducing |
|---|
| **Step I** |
| For each lane do: |
|     For each original position $O_c$ do: |
|       utilize IDM to infer the complete trajectory before queue |
|     For each collected destination position $D_c$ do: |
|       search for the destinations $D_i$ inferred from IDM during the range of [-5s, +5s] |
|       selected a nearest $D_i$ as the destination of un-lane change trajectory |
|       the remaining $O_c$, $D_c$ and $D_i$ might be regarded as the LCV |
| **Step II** |
| For All lanes do: |
|     sort all $O_c$ and $D_c$ by timestamp |
|     For each collected destination position $D_c$ do: |
| search for the $D_i$ during the range of [-5s, +5s] in adjacent lanes |
| selected a nearest $D_i$ as the destination of lane change trajectory, and determined the target lane |
| the remaining $O_c$, $D_c$ and $D_i$ might be influenced by Queue |
| **Step III** |
| For each lane do: |
| count the remaining original position $O_c$ and destination position $D_c$ |
| if $O_c == D_c$: |
| All trajectories will regard as Non-LCVs，and match in order |
| elif $O_c < D_c$: |
| The vehicles of $O_c$ are No-LCV, match with $D_c$ in order |
| else: |
| The vehicles of $D_c$ are No-LCV, match with $O_c$ in order |
| the remaining $O_c$ and $D_c$ are regarded LCV |
| **Step IV** |
| For All lanes do: |
| sort all remaining $O_c$ and $D_c$ by timestamp |
| For each $O_c$ do: |
| search for the closest $D_c$ from adjacent lanes as the LCV, and determined the target lane |

## 3.2. Lane Change Behavior Module (LC-GAN)

The LC process comprises two distinct phases: decision-making and execution, with the latter typically lasting 2-3 seconds. Given the urban intersection speed limit of 40 km/h, this duration corresponds to a longitudinal displacement of 20-30 meters during LC behavior. To precisely localize LC positions while accounting for vehicle dimensions (average length of 4.5 m) and minimum inter-vehicle spacing (1.5 m), we discretize the study area into spatiotemporal blocks for each specific vehicle with a resolution of 6 m. The resulting block sequence, denoted as $S$ in Fig. 2, employs binary encoding, where '1' marks the block containing the LC event and '0' represents all other blocks. For instance, as illustrated in Fig. 2, in a 60 m study section (divided into 10 blocks), an LC occurring between 24 m and 30 m from the origin activate the 5-th block and set its value to 1. This discrete block representation serves as the output format for the LC-GAN model and offers several advantages, including quantizing continuous space into computationally manageable discrete units, maintaining essential spatial relationships through sequence position, and enabling direct integration with neural network architectures.

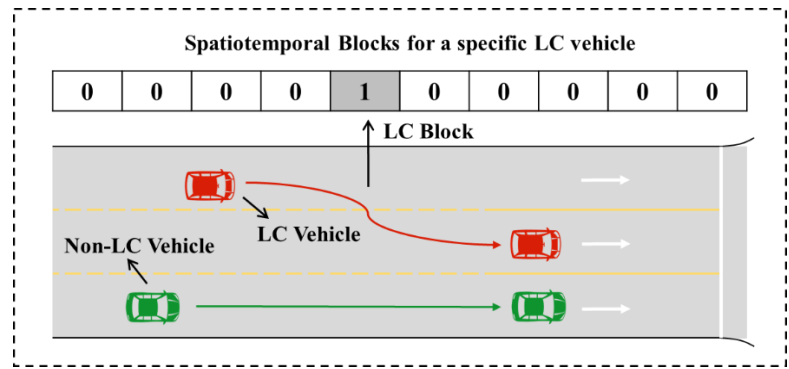

Fig. 2 Schematic diagram of spatiotemporal blocks for a specific vehicle

Mengyun Xu, Jie Fang, Eui-Jin Kim, Tony Z. Qiu, Prateek Bansal

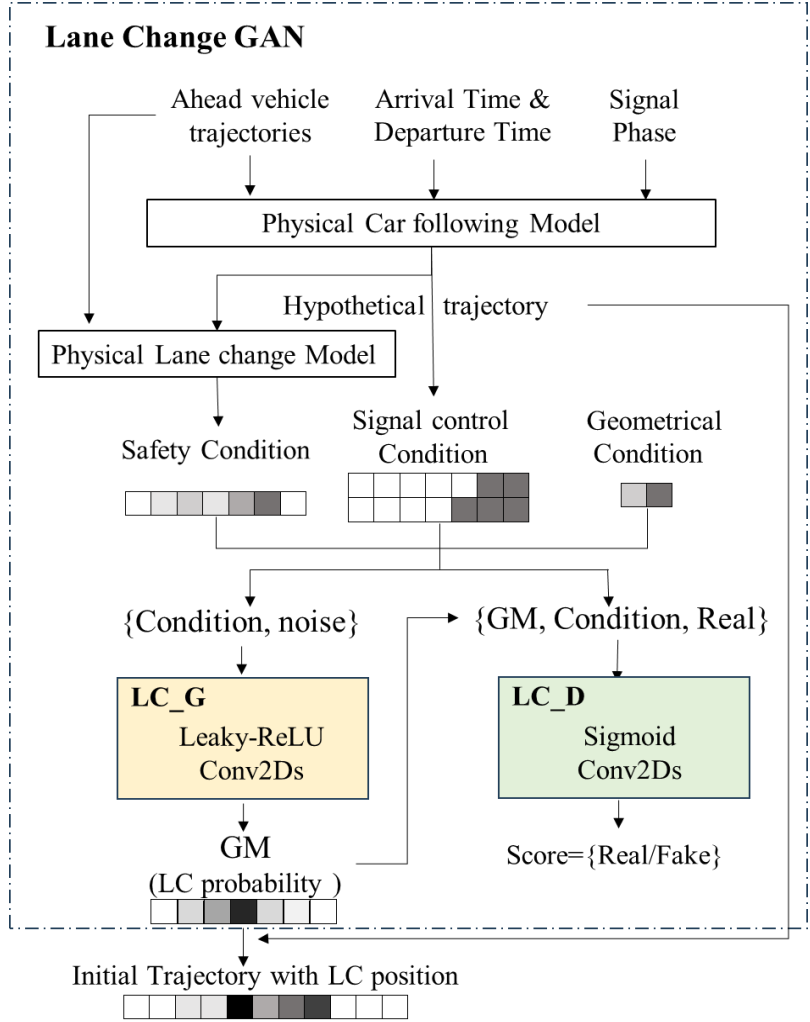

Fig. 3 The architecture of the Lane Change GAN

Fig. 3 illustrates the architecture of the *Lane Change Behavior Module*, which processes three primary inputs: AVTs, arrival and departure timestamps, and signal phase information. The module first generates complete hypothetical trajectories for both origin and destination lanes using a physical car-following model (e.g., IDM), thereby simulating continuous vehicle movement from the entry to exit points within the study section. These hypothetical trajectories represent physically plausible yet mutually exclusive motion paths, as a vehicle cannot simultaneously occupy two lanes.

Subsequently, a physical LC model (e.g., MOBIL) is applied to derive critical LC-related information, including safety-aware LC probabilities influenced by surrounding vehicle interactions, signal control, and geometric configuration. This physics-derived information serves as conditioning inputs for the cGAN architecture. Through adversarial training, the generator (**LC_G**) and discriminator networks (**LC_D**) iteratively optimize against each other to produce precise LC probability distribution that align with physical and behavioral realism.

Finally, the module outputs a complete initial trajectory with accurately localized LC positions by combining the relevant sub-trajectories, segmented at the maximum-probability LC block, from both the current and target lane hypothetical trajectories. The specific details of each subcomponent are described in the following subsections.

### 3.2.1 Physical Car Following Model

A well-established car-following model IDM is adopted for generating hypothetical trajectories in this study. The acceleration of the Lag HV is determined by the vehicle's current speed, the relative speed with respect to the leading vehicle (e.g., CV), and the gap between the two vehicles, as expressed in Equations 2 and 3. The subsequent kinematic state of the lag vehicle at next timestamp $t+1$, velocity $v_{\text{lag}}^{t+1}$ and the position $x_{\text{lag}}^{t+1}$, are then is propagated through Equations 4 and 5:

$$\dot{v}_{\text{lag}}^{t} = a\left[1 - \left(\frac{v_{\text{lag}}^{t}}{dv_{\text{lag}}^{0}}\right)^{\delta} - \left(\frac{s^{*}(v_{\text{lag}}^{t}, v_{\text{cv}}^{t})}{s_{\text{lag}-\text{cv}}^{t}}\right)^{2}\right] \qquad (2)$$

$$s^{*}(v_{\text{lag}}^{t}, v_{\text{cv}}^{t}) = s_0 + v_{\text{lag}}^{t}T + \frac{v_{\text{lag}}^{t}(v_{\text{lag}}^{t} - v_{\text{cv}}^{t})}{2\sqrt{ab}} \qquad (3)$$

$$v_{\text{lag}}^{t+1} = v_{\text{lag}}^{t} + \dot{v}_{\text{lag}}^{t} \qquad (4)$$

$$x_{\text{lag}}^{t+1} = x_{\text{lag}}^{t} + v_{\text{lag}}^{t} + \frac{1}{2}\dot{v}_{\text{lag}}^{t} \qquad (5)$$

where $cv$ denotes the leading CV with a known trajectory and $lag$ denotes the following HV whose trajectory needs to be reconstructed. $\dot{v}_{\text{lag}}^{t}$ denotes acceleration of lag vehicle in time stamp $t$. $v_{\text{lag}}^{t}$ and $v_{\text{cv}}^{t}$ are the velocity of lag and leading vehicles at timestamp $t$, respectively. $dv_{\text{lag}}^{0}$ is the desired speed of lag vehicle, which is calculated by averaging the detected CV velocity and the historical segment running speed. $T$ is the desired time headway. $s_{\text{lag}-\text{cv}}^{t}$ is the instantaneous vehicle spacing in between lag vehicle and its leading vehicle. $\delta$ is the acceleration parameter, typically set to 4 in arterial traffic. $s^{*}(v_{\text{lag}}^{t}, v_{\text{cv}}^{t})$ is the desired following gap, consisting of the equilibrium term $(s_0 + v_{\text{lag}}^{t}T)$ and the dynamic term $\left(\frac{v_{\text{lag}}(v_{\text{lag}} - v_{\text{cv}})}{2\sqrt{ab}}\right)$ for implementing the "intelligent" braking strategy. $s_0$, $a$ and $b$ are the parameters for desired minimum gap to the front vehicle, the maximum acceleration, and the comfortable deceleration, respectively. These IDM parameters can be calibrated to reflect various driving behaviors and local traffic conditions (Treiber et al., 2000).

The detector-based vehicle arrival timestamps are used to initialize each vehicle trajectory at the segment entrance ($x_{\text{lag}}^{t=1}$). The following gap $s_{\text{lag}-\text{cv}}^{t}$ is directly obtained from detector observations. The initial velocity $v_{\text{lag}}^{t=1}$ of the lag vehicle is assumed equal to that of the leading CV ($v_{\text{cv}}^{t=1}$). Then, the subsequent vehicle state is recursively deduced using Equations 2 to 5 until the position $x_{\text{lag}}^{t}$ reaches the end of the queue. The position of joining the queue can be inferred by adding the empirically observed average inter-vehicle spacing.

### 3.2.2 Physical lane change model informed Condition information

Existing LC models predominantly focus on traffic efficiency, safety, and driving comfort. For instance, the Safe LC Domain approach establishes collision-free regions for LC while minimizing path curvature to enhance ride comfort. The Minimizing Overall Braking Induced by Lane changes (MOBIL) model, widely implemented in SUMO simulations, reduce system-wide braking by optimizing acceleration benefit. However, arterial environments differ fundamentally from freeways due to shorter inter-intersection distances, stronger signal influences, and smaller velocity or acceleration gains from LCs. Consequently, LC modeling for arterial environments requires distinct behavioral considerations.

Specifically, for MLC, drivers tend to execute early LCs within the limited roadway segment while maintaining safety margins. In contrast, DLC exhibit higher behavioral randomness driven by traffic signal timing: during green phases, drivers accelerate to pass the intersection by exploiting available gap, whereas during red phases, they prefer shorter queues to minimize expected delay, often initiating LCs near the queue tail. To reflect these arterial-specific behaviors, the LC-GAN incorporates three physical conditions:

- Safety condition $C_{safety}$

The safety condition $C_{safety}$ represent the influence of surrounding vehicles in determining feasible LC positions. Imposing a safety-gap condition effectively prevent physically infeasible or collision-prone maneuvers. Across each spatiotemporal block, the spatial gap $gs$ between the inferred hypothetical trajectory of LCV and its possible preceding vehicle in the target lane is evaluated. Following Ahmed (1999), we optimize the safety-aware utility function to compute the LC feasibility $P_{safety}$ for each block, as shown in Equation 6. If $gs$ exceeds the acceptable gap $\varphi_1$, $P_{safety} = 1$; if $gs$ fall below the critical gap $\varphi_2$, $P_{safety} = 0$, indicating imminent collision risk. For intermediate gaps, the classical physical logit model yields the safety-aware LC feasibility as:

$$P_{safety} = \begin{array}{ll} 1 & gs \geq \varphi_1 \\ \frac{1}{1+e^{-(\beta_0 + \beta_1 \cdot [gs - (\varphi_2 - \varphi_1)])}} & \varphi_1 \leq gs \leq \varphi_2 \\ 0 & gs \leq \varphi_2 \end{array} \qquad (6)$$

where $\beta_0$ is a constant and $\beta_1$ quantifies sensitivity to the safety margin. When $P_{safety}$ exceeds zero, it indicates that a LC behavior is feasible within the corresponding spatiotemporal blocks. Furthermore, as $P_{safety}$ approaches 1, the feasibility of executing the LC increases correspondingly. Embedding $C_{safety}$ into the LC-GAN confines LCs to blocks where $P_{safety} > 0$, thereby eliminating unsafe maneuvers.

- Signal control condition $C_{signal}$

Traffic signal control leads to heterogeneous queue length distributions across lanes, which substantially influence LC decisions. The queue lengths of the current and target lanes exert a decisive impact on LC decisions, as drivers tend to select shorter queues to minimize departure delays at downstream signalized intersections. This behavioral mechanism is formally incorporated into our framework as a signal control condition ($C_{signal}$), representing the influence of signal control on LC feasibility.

Considering traffic regulatory conditions, this study assumes that once a vehicle joins the tail of queue in its original lane, no further LCs are permitted. LCs are therefore allowed only before queue formation or during queue dissipation in the target lane. To capture these dynamics, the time-varying queue states on both the origin $C_{signal_o}$ and target $C_{signal_t}$ lanes are incorporated as critical conditional inputs.

Formally, the LC feasibility is then implemented by analyzing the stopping state of the preceding vehicle at each spatiotemporal block along the hypothetical trajectory. As shown in Fig. 4, this framework assigns a LC feasibility of 1 to blocks preceding queue formation, 0 to blocks within stationary queues, and 1 to part of blocks during queue dissipation where vehicle movement resumes and LC become physically feasible.

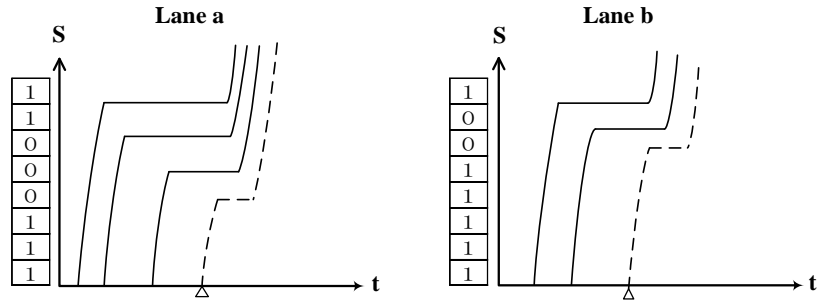

Fig. 4 the example of signal control condition corresponding with queue length

- Geometrical condition $C_{geo}$

Different combination of original and target lanes reflects distinct LC decisions. As illustrated in Fig. 5, a transition from lane **b** to lane **a** typically corresponds to left-turning vehicles, representing MLC, whereas transitions between **b** and **c** (both through lanes) indicate DLC. A transition from lane **c** to lane **a** means a secondary LC event. Furthermore, the spatial distribution of LC positions in DLC varies with roadway geometry: for example, upstream through-traffic typically enter lane **b** first, while upstream right-turning vehicles might enter lane **c** first, resulting in different LC position distributions across lane pairs.

To effectively capture these lane-pair-specific geometrical characteristics, lane-pair identified are encoded as a geometrical condition ($C_{geo}$), which categorizes LC types based on roadway geometric configuration. As shown in the Fig. 6, each LC behavior is represented by a unique origin-target (O-T) lane-pair vector of size 1×2. A one-hot encoding is applied to all O-T lane combinations, followed by an embedding transformation that generate 1×S feature sequence. Subsequently, $C_{geo}$ is incorporated into the cGAN framework along with other physical conditions.

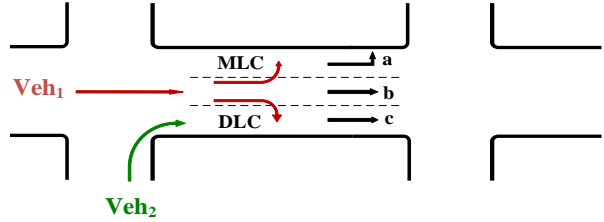

Fig. 5 the example of different LC intentions, including MLC and DLC

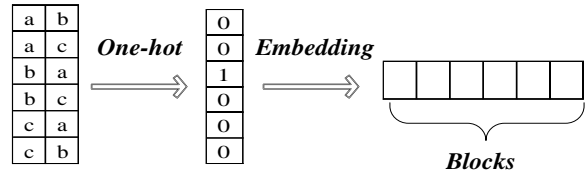

Fig. 6 the example of geometrical condition

### 3.2.3 Lane Change GAN

The operation of traffic signals induces heterogeneous vehicle arrival patterns at downstream intersections, exhibiting platoon-dominated arrivals during initial green phases and progressively dispersed arrivals in subsequent intervals. Preliminary data analysis reveals that vehicles arriving in dispersed patterns demonstrate significantly more scattered LC-probability distributions than those in platoon formations. Driver heterogeneity further contributes to this variability: aggressive drivers may execute last-minute LC near queue tails, even when sufficient safety margins exist upstream. In contrast, conservative drivers tend to initiate LC early at movement origins to ensure both comfort and successful execution.

Given these complexities, traditional parametric models (e.g., logit model) face inherent limitations in accurately estimating LC position. Their rigid distributional assumptions cannot simultaneously accommodate phase-dependent arrival patterns, heterogeneous driving behaviors, and the intrinsic differences between MLC and DLC mechanisms. In contrast, GANs offer a flexible, data-driven framework for modeling such nonlinear and multimodal distributions, particularly suited for LC position estimation in arterial environments. Specifically, adversarial training of GAN enables the generator to dynamically approximate complex real-world distributions without any predefined functional assumptions, while naturally accommodating both stochasticity and structural diversity in arterial LC behaviors.

To estimate LCPs conditioned on relevant information, we employ a physics-informed cGAN structure, hereafter referred to as the Lane Change GAN (LC-GAN). The primary idea behind cGAN is that the model estimates the distribution of observed data by leveraging both physical conditions (i.e., signal control, safety, and geometrical conditions) and random latent noise. LC-GAN thus estimates the distribution of LC positions across heterogeneous vehicles without prior parametric assumptions on LC probability distributions.

As shown in Fig. 3, LC-GAN comprises a generator **LC_G** and a discriminator **LC_D**, with detailed components described in the following subsections. The generator **LC_G** aims to learn a mapping function from the arbitrary input noise vector to the target data distribution (i.e., the LC probability distributions) reflecting to efficiency and safety features inferred from the physical lane change model. Accordingly, the input to **LC_G** includes a sequence of random noise $Z_{lc}$ and multiple physical condition sequences $C_{lc}$. The lengths of $Z_{lc}$ and $C_{lc}$ are both set equal to the number of discretized spatiotemporal blocks. The noise sequence $Z_{lc}$ is sampled from a predefined probability distributions (e.g., Gaussian distribution), while the physical condition input $C_{lc}$ is designed to account for the influence of the safety-based LC feasibility $C_{safety}$, signal control correlated with queue length in original $C_{signal_o}$ and target $C_{signal_t}$ lanes, and the geometric lane identifiers $C_{geo}$. The detailed composition of these physical condition input $C_{lc}$ is described in Section 3.3.2. Four physical conditions are concatenated and fed into **LC_G** and **LC_D** as follows:

$$C_{lc} = \{C_{safety} \parallel C_{signal_o} \parallel C_{signal_t} \parallel C_{geo}\} \tag{7}$$

Subsequently, $C_{lc}$ is concatenated with $Z_{lc}$ to establish a mapping from the noise distribution of $Z_{lc}$ to the LC probability distribution $GM_{lc}$.

$$GM_{lc} = \textbf{LC\_G}(Z_{lc}, C_{lc}) \tag{8}$$

The generator **LC_G** adopts multi-layer convolutional architecture with batch normalization to hierarchically extract and integrate spatiotemporal features. This design effectively model complex, high-dimensional interactions underlying LC position dynamics. The output layer of **LC_G** employs a *softmax* activation function to generate a probability distribution across spatiotemporal blocks. During inference, we apply *argmax* operation to select the block with highest LC probability.

The discriminator **LC_D** aims to determine whether a given LC probability distribution is real or fake, producing a scalar $score_{lc}$ that reflects how well the input align with its associated conditional information $C_{lc}$. The input consists of (i) the distribution of LC probability in the field of experiment, which can be either the real sequence $RM_{lc}$ or the generated sequence $GM_{lc}$, and (ii) a physical condition sequence $C_{lc}$.

$$score_{lc} = \textbf{LC\_D}(GM_{lc}, RM_{lc}, C_{lc}) \tag{9}$$

In the structure of **LC_D**, a convolution layer followed by Leaky-ReLU layer is applied to explore the correlations among combined input features in discriminator, as presented in Fig. 3. After that, multiple convolutional layers enable **LC_D** to model latent relationships and classify the input as real or fake. **LC_D** increases the output score when the input corresponds to a real LC distribution and decreases it when the input is generated. Moreover, consistent with cGAN structure, **LC_D** is also designed to assign low scores to mismatched combination of $RM_{lc}$ and shuffled condition sequences, denoted as $SC_{lc}$. By back propagating the gradients from the **LC_D** to the **LC_G**, the generator is guided to produce LC probability distributions that are both physically consistent and behaviorally realistic.

LC-GAN adopts a combination of reconstruction and adversarial losses to jointly enhance the realism and accuracy of the generated LC probability distributions. During adversarial training, the generator **LC_G** is trained to maximize the likelihood that the discriminator **LC_D** misclassifies generated samples as real, while **LC_D** is trained to maximize its ability to correctly distinguish real samples from the generated ones. In addition to adversarial loss, we incorporate a reconstruction loss (i.e., MSE or cross-entropy) between the real and the generated LC probability sequences to promote stable and efficient training. This combined loss function allows the LC-GAN to maintain high fidelity in reproducing the underlying distribution while preserving the physical and behavioral consistency of the generated LC patterns.

The loss functions for **LC_G** and **LC_D** are shown in Equations 10 to 13. The adversarial loss term $\mathcal{L}_{lcD1}$ for **LC_D** is designed to assign a higher score to the real distribution. The reconstruction loss term $\mathcal{L}_{lcD2}$ minimizes the MSE between the real and generated sequence values. Meanwhile, the generator loss term $\mathcal{L}_{lcG}$ is formulated to encourages the discriminator **LC_D** to assign higher scores to generated distributions, guiding **LC_G** to produce output that resemble real distribution.

$$\mathcal{L}_{lcD1} = \frac{1}{B}\{\log\big(1 - \textbf{LC\_D}(GM_{lc}, C_{lc})\big) + \log\big(1 - \textbf{LC\_D}(RM_{lc}, SC_{lc})\big) + \log\big(\textbf{LC\_D}(RM_{lc}, C_{lc})\big)\} \tag{10}$$

$$\mathcal{L}_{lcD2} = -\frac{1}{B}(RM_{lc} - \textbf{LC\_G}(Z_{lc}, C_{lc}))^2 \tag{11}$$

$$\mathcal{L}_{lcD} = \mathcal{L}_{lcD1} + \omega_{lc}\mathcal{L}_{lcD2} \tag{12}$$

$$\mathcal{L}_{lcG} = -\frac{1}{B}log\big(\textbf{LC\_D}(GM_{lc}, C_{lc})\big) \tag{13}$$

where $\mathcal{L}_{lcG}$ and $\mathcal{L}_{lcD}$ are the loss functions of the generator **G** and discriminator **D**, respectively. $B$ is the batch size. $\omega_{lc}$ represents a weighting coefficient that balances the relative contributions of the adversarial loss term $\mathcal{L}_{lcD1}$ and reconstruction loss term $\mathcal{L}_{lcD2}$ to $\mathcal{L}_{lcD}$.

### 3.3. Trajectory Reconstruction Module for different lanes (Trajectories GAN)

Building upon previous work (Xu et al., 2025), vehicle trajectories are reconstructed using a physics-informed Trajectory-GAN model. The proposed model generates behaviorally heterogeneous and traffic-adaptive trajectories through a two-stage process. First, initial trajectories estimated by physics-based car-following models are treated as physical conditions. Second, the Trajectory-GAN model refines these initial trajectories within a conditional GAN framework to capture unobserved car-following behaviors, enabling the reconstructed vehicle trajectories to dynamically adjust to varying traffic conditions. While reconstructing trajectories across multiple lanes, as illustrated in Fig. 7, the framework employs distinct reconstruction mechanisms for LC and non-LC vehicles to accurately reflect their differing motion dynamics and behavioral patterns.

For LC vehicles, hypothetical trajectories from both origin and target lanes are first generated using the physical lane change model and then integrated with the LC position predicted by the LC-GAN to form a new initial trajectory. For non-LC vehicles, initial trajectories are inferred directly from physics-based car-following model, conditioned on the AVTs, arrival and departure time, and signal phase information.

In addition to using the initial trajectory as the conditional sequence $C_{traj}$, we also input a sequence of noise $Z_{traj}$ into the generator **Traj_G** for incorporating stochastic driving characteristics. **Traj_G** adopts a Bidirectional Long Short-Term Memory (Bi-LSTM) architecture to capture the intricate temporal and spatial dependencies within trajectories. This structure enable the neural network to exploit sequence information in both forward (past-to-future) and reverse (future-to-past) directions, allowing us to effectively model the complex spatiotemporal relationships inherent in each trajectory point. The output of Bi-LSTM is fed into a convolutional layer to extract high-dimensional features to explore the underlying spatial-temporal interactions. Subsequently, the batch normalization and activation function (leaky-ReLU) are attached to generate the final output.

$$GM_{traj} = \textbf{Traj\_G}(Z_{traj}, C_{traj}) \tag{14}$$

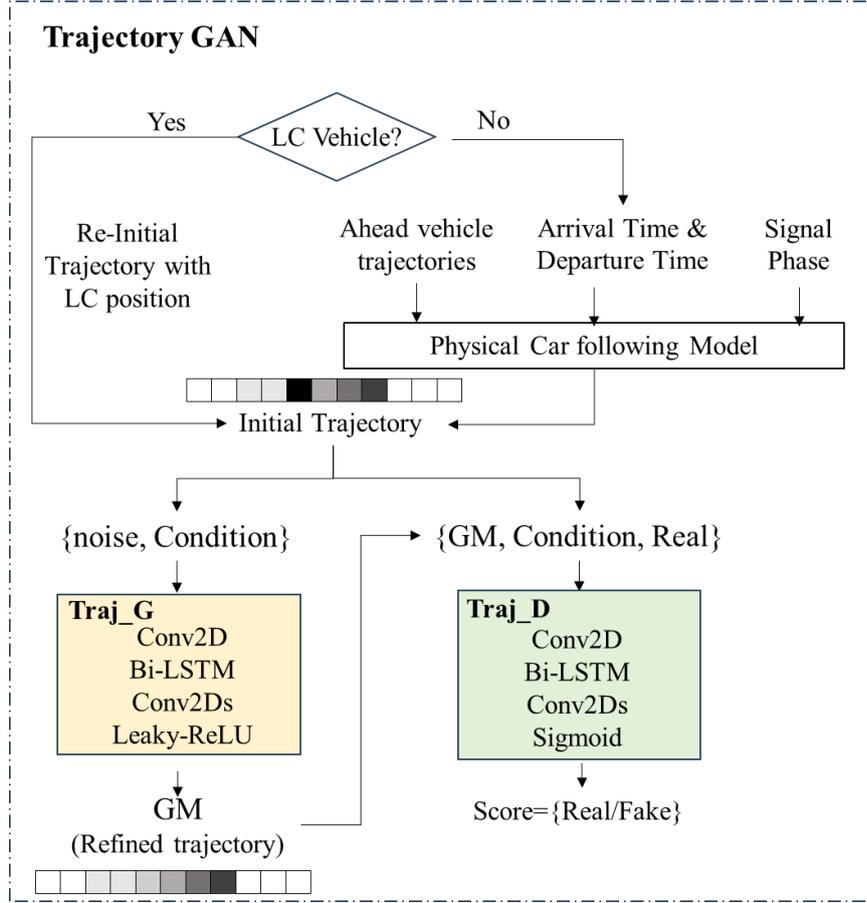

Fig. 7 The architecture of the Trajectory GAN

The discriminator **Traj_D** in the Trajectories-GAN takes two inputs: i) a generated trajectory $GM_{traj}$ or a collected real trajectory $RM_{traj}$, representing the sequence of vehicle position over one signal cycle, and ii) a condition sequence $C_{traj}$. The output of **Traj_D** is a scalar value indicating whether the input trajectory is real or generated, as well as whether the pairing between $RM_{traj}$ and $C_{traj}$ is consistent. The intricate structure of **Traj_D** closely mirrors that of **Traj_G**. Empirical investigations have demonstrated that the adoption of analogous architectural configurations for both generator and discriminator facilitates stable model convergence and enhance overall reconstruction accuracy.

$$score_{traj} = \textbf{Traj\_D}(GM_{traj}, RM_{traj}, C_{traj}) \tag{15}$$

For the training process, we employ an adversarial training procedure that integrates adversarial loss and reconstruction loss (MSE). $\mathcal{L}_{TrajD}$ and $\mathcal{L}_{TrajG}$ denote loss functions for **Traj_D** and **Traj_G**, respectively. By jointly optimizing **Traj_G** and **Traj_D** under the dual objectives of adversarial and reconstruction losses, the framework achieves more stable convergence and improved reconstructed performance.

$$\mathcal{L}_{TrajD1} = \frac{1}{B}\{\log\left(1 - \textbf{Traj\_D}\left(GM_{traj}, C_{traj}\right)\right) + \log\left(1 - \textbf{Traj\_D}\left(RM_{traj}, SC_{traj}\right)\right) + \log\left(\textbf{Traj\_D}\left(RM_{traj}, C_{traj}\right)\right)\}$$

$$\tag{16}$$

$$\mathcal{L}_{TrajD2} = -\frac{1}{B}\sum_{c=1}^{B}\left(RM_{traj} - \textbf{Traj\_G}\left(Z_{traj}, C_{traj}\right)\right)^2 \tag{17}$$

$$\mathcal{L}_{TrajD} = \mathcal{L}_{TrajD1} + \omega_{traj}\mathcal{L}_{TrajD2} \tag{18}$$

$$\mathcal{L}_{TrajG} = -\frac{1}{B} log(\boldsymbol{D}(GM_{traj}, C_{traj})) \qquad (19)$$

where $SC_{traj}$ is a mismatched conditional sequence. $B$ is the batch size, which is equal to the number of trajectories in our dataset. $\omega_{traj}$ represents a weighting coefficient that balances the relative contributions of the adversarial loss term $\mathcal{L}_{TrajD1}$ and reconstruction loss term $\mathcal{L}_{TrajD2}$ to $\mathcal{L}_{TrajD}$.

## 3.4. The multi-task joint generative learning

As illustrated in Fig. 1, the LC behavior generated by the LC GAN module directly affects the initial trajectory inputs for the Trajectory-GAN, while the optimized results from the Trajectory-GAN simultaneously constrains the estimation of LC positions. To model this bidirectional dependency, we designed a multi-task joint generative learning that establishes a synergistic interaction between LC behavior modeling and trajectory reconstruction, allowing both components to mutually inform and regularize each other.

Specifically, our method jointly optimizes the loss functions of both the LC-GAN and Trajectory-GAN within a unified training framework, as provided in the **Algorithm2** pseudocode below. This multi-task joint generative learning enables the simultaneous training of both GANs, thereby addressing the objectives of two related tasks in a single coherent process. Through this co-training design, the two GANs exchange informative cues, learn shared spatiotemporal representations, and capture latent correlations between LC behavior and trajectory construction, ultimately enhancing overall model performance. Consequently, this co-optimization framework enhances the realism and reliability of multi-lane vehicle trajectory reconstruction through two key mechanisms: (1) ensuring spatiotemporal consistency between LC behaviors and trajectory reconstructions, and (2) maintaining physical plausibility across all generated trajectories.

| **Algorithm 2: Joint training process of LC GAN and Traj-GAN** |
| --- |
| **Input:** |
| Random Noise vectors: $Z_{lc}$, $Z_{traj}$ |
| Real data: $RM_{lc}$, $RM_{traj}$ |
| Other inputs: CV trajectory, Ahead vehicle trajectories, Arrival Time, Departure Time, Signal Phase |
| **Output:** |
| Trained generators: **LC_G** , **Traj_G** |
|  |
| 1:     for each training iteration do |
|       //step1: Update Discriminators (fixed **G**) |
| 2:     Hypothetical trajectories = physical car following model (CV trajectory, Arrival Time, Departure Time, Signal Phase) |
| 3:     $C_{lc}$ = physical lane change model (Ahead vehicle trajectories, Hypothetical trajectories) |
| 4:     $GM_{lc}$= **LC_G** $(Z_{lc}, C_{lc})$       // generated LC position |
| 5:     $C_{traj}$ = Trajectory reorganize (Hypothetical trajectories, $GM_{lc}$)       // initial trajectory with LC position |
| 6:     $GM_{traj}$= **Traj_G**$(Z_{traj}, C_{traj})$       // refined trajectory |
| 7:     $\mathcal{L}_{lcD}$ calculated from equation 10 to 12 |
| 8:     $\mathcal{L}_{TrajD}$ calculated from equation 16 to 18 |
| 9:     $\mathcal{L}_D = \mathcal{L}_{lcD} + \mathcal{L}_{TrajD}$ |
| 10:    $\mathcal{L}_D$.backward()       // Backprop through both **LC_D** and **Traj_D** |
| 11:    **D**.step()       // Update both $D_1$ and $D_2$ parameters simultaneously |
|       // step 2: Update Generators (fixed **D**) |

| 12: | $GM_{lc}=$ **LC_G** $(Z_{lc}, \ C_{lc})$ |
|---|---|
| 13: | $C_{traj}$ = Trajectory reorganize (Hypothetical trajectories, $GM_{lc}$) |
| 14: | $GM_{traj}=$ **Traj_G**$(Z_{traj}, \ C_{traj})$ |
| 15: | $\mathcal{L}_{lcG}$ calculated from equation 13 |
| 16: | $\mathcal{L}_{TrajG}$ calculated from equation 19 |
| 17: | $\mathcal{L}_G= \mathcal{L}_{lcG} \ + \mathcal{L}_{TrajG}$ |
| 18: | $\mathcal{L}_G$.backward()　　　　// Backprop through both **LC_G** and **Traj_G** |
| 19: | **G**.step()　　　　// Update both $G_1$ and $G_2$ parameters simultaneously |
| 20: | end for |

The joint loss functions for the discriminator and generator networks are mathematically formulated in Lines 9 and 17 in the **Algorithm 2**. In step 1, we simultaneously update the parameters of both **LC_D** and **Traj_D** while maintaining fixed generator parameters; In the subsequent step 2, the gradient of the loss function of **D** is fed back to **G** to optimize the parameter of both **LC_G** and **Traj_G**. This iterative adversarial training process, where the **G** (comprising **LC_G** and **Traj_G**) and **D** (consisting of **LC_D** and **Traj_D**) networks are alternately updated, continues until convergence is achieved, as indicated by the stabilization of loss values across successive iterations. To mitigate the risk of gradient explosion during multiple GAN training, gradient normalization is applied throughout the joint learning process.

## *4.* Empirical results

### 4.1. Experimental Design and Evaluation Indicators

We empirically validate the proposed framework using two publicly available real-world datasets: the DRone-derived Intelligence for Traffic analysis (DRIFT) dataset from South Korea and the Next Generation Simulation (NGSIM) dataset from the United States.

The DRIFT dataset, an open-source resource published in 2025, provides high-resolution spatiotemporal trajectories of all vehicles within urban intersections, captured using unmanned aerial vehicle (UAV) platforms. This dataset offers high-quality ground-truth information for validating microscopic traffic flow models and trajectory reconstruction algorithms. To ensure sufficient representation of LC events, we focus on the eastbound approach of *Intersection F*, a signalized intersection with frequent LC behavior.

The intersection layout and lane configuration are depicted in Fig. 8, which consists of one exclusive left-turn lane and two through lanes. The spatial distribution of LCPs within the study area is presented in the right-hand panel of Fig. 8. The trajectory reconstruction covers a 200-meter section upstream of the stop line, which is divided into 34 spatiotemporal blocks (= 200 m / 6 m resolution). The dataset includes 941 vehicle trajectories, among which 293 exhibit LC behaviors. For evaluation, trajectory data from two complete signal cycles (approximately 20% of the total sample) are allocated for testing, while the remaining data are used for model training and validation.

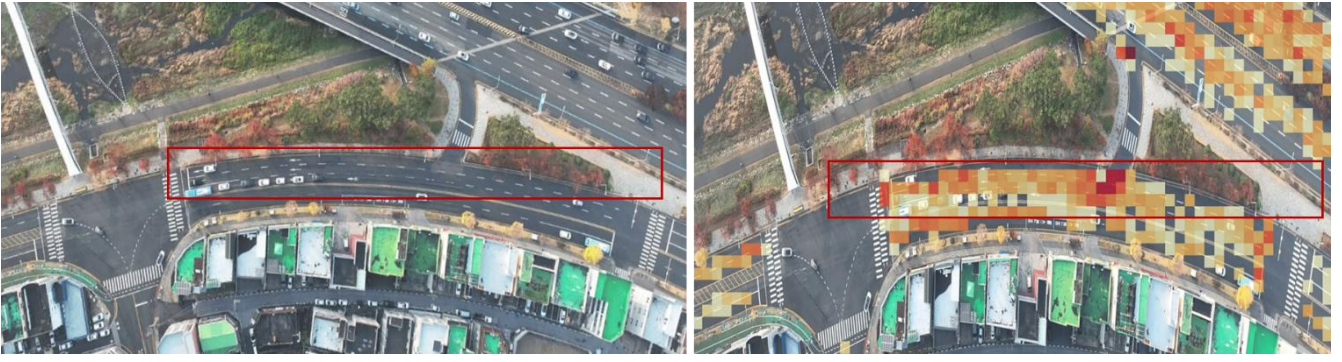

Fig. 8 Layout of the studied area and the spatial distribution of lane change positions in the DRIFT

To further assess the adaptability of the proposed trajectory reconstruction framework across different urban contexts, additional validation was conducted using the widely cited NGSIM dataset (Montanino and Punzo, 2013). As illustrated in Fig. 9, the study area covers the westbound approach of the intersection at Lankershim St. and Universal Hollywood Dr in Los Angeles, California. This approach consists of five lanes: one exclusive right-turn lane, three through lanes, and one exclusive left-turn lane, with detailed lane markings provided in Fig. 9.

The reconstruction covers 260 feet upstream segment from the stop line, divided into 13 contiguous blocks, each spanning 20 feet (approximately 6 m resolution). The dataset contains 813 vehicle trajectories, of which 203 exhibit LC behaviors. The trajectory data was partitioned as 80% for training, 10% for validation, and 10% for testing. Testing samples were selected from a complete signal cycle to ensure temporal consistency.

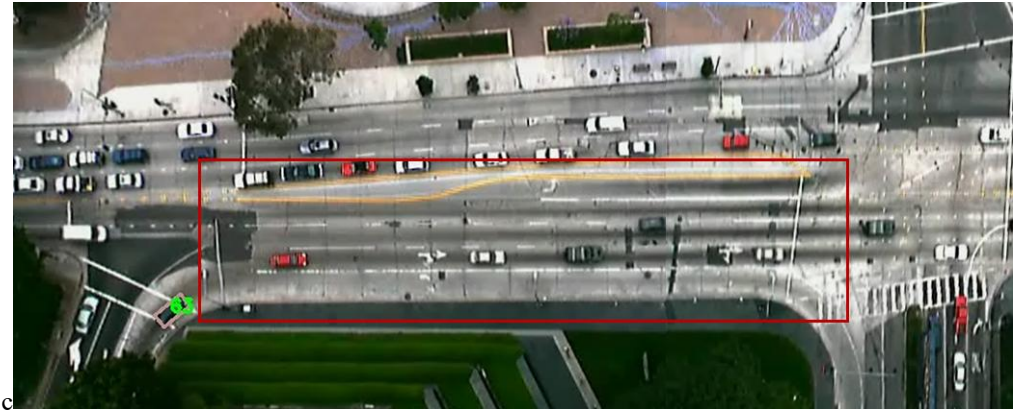

Fig. 9 Layout of the studied area in the NGSIM

From a microscopic perspective, we evaluate the performance of reconstructed trajectory using three indicators: Queuing Error (QE) for queuing status, and Time Error (TE) and Position Error (PE) for moving status. The specific formula can be found in Equations 20 to 22.

$$QE = \sqrt{\frac{1}{QT} \sum_{qt=1}^{QT} (P_{qt}^{\text{real}} - P_{qt}^{\text{rec}})^2} \tag{20}$$

$$TE = \sqrt{\frac{1}{MP} \sum_{mp=1}^{MP} (T_{mp}^{\text{real}} - T_{mp}^{\text{rec}})^2} \tag{21}$$

$$PE = \sqrt{\frac{1}{MT} \sum_{mt=1}^{MT} (P_{mt}^{\text{real}} - P_{mt}^{\text{rec}})^2} \tag{22}$$

where $QT$ and $MT$ represent the number of temporal evaluation points (i.e., time duration) during the queuing status and moving status, respectively. For $QE$, $qt \in \{1, \dots, QT\}$ denotes discrete time indices within the queuing period, and $P_{qt}^{\text{real}}$ and $P_{qt}^{\text{rec}}$ are the ground-truth and reconstructed vehicle positions at the same time step $qt$. For $PE$, $mt \in \{1, \dots, MT\}$ denotes discrete time indices during the moving status, and $P_{mt}^{\text{real}}$ and $P_{mt}^{\text{rec}}$ are the ground-truth and reconstructed vehicle positions at the same time step $mt$. For $TE$, $MP$ denotes the number of spatial evaluation points used during the moving status, and $mp \in \{1, \dots, MP\}$ represents the spatial index. $T_{mp}^{\text{real}}$ and $T_{mp}^{\text{rec}}$ are the time stamps at which the ground-truth and reconstructed trajectories, respectively, pass the same spatial position $mp$.

The estimation accuracy of LCPs is a critical factor influencing the precision of reconstructed trajectories. In this study, LCPs are represented using discretized spatiotemporal blocks. We therefore employ the Block Error (BE) metric to evaluate LC estimation performance for each vehicle, defined as:

$$BE = \frac{1}{N}\sum_{n=1}^{N}\left|Block_n^{\text{real}} - Blocl_n^{\text{est}}\right| \tag{23}$$

where $Block_n^{\text{real}}$ and $Blocl_n^{\text{est}}$ denotes the ground-truth and estimated block indices of LCP for vehicle $n$, respectively. A smaller BE value indicates higher accuracy in predicting the precise spatiotemporal location of lane-change events.

## 4.2. Performance Evaluation of LC GAN

### 4.2.1 Performance comparison with other LC Models

The proposed framework employs the LC-GAN to estimate LC probabilities across discretized spatiotemporal blocks, selecting the block with the highest probability as the predicted LC position. The corresponding Block Error (BE), which quantifies the deviation between predicted and ground-truth LCPs, is summarized in Table 1.

Within the testing cycles of the DRIFT dataset, 33% of LCVs performed DLC, while 67% executed MLC. Similarly, in the NGSIM testing cycles, 40% of LC were classified as DLC and 60% as MLC. To validate the effectiveness of the proposed LC-GAN, three baseline models were considered for comparison:

1) Rule-based LC model, which performs an LC whenever an acceptable safety gap is available;

2) Utility-based LC model, which integrates speed utility and safety criteria, based on the classical Minimizing Overall Braking Induced by Lane Changes (MOBIL) framework (Kesting et al., 2007);

3) Data-driven LC model without physical conditions, which employ deep learning architectures, specifically Deep Belief Network (DBN) and LSTM networks, to model both LC decisions and execution (Xie et al., 2019).

Table 1. The block errors in different baselines across various type of LC

| Block Error | DRIFT(34blocks) | | | NGSIM(13blocks) | | |
|---|---|---|---|---|---|---|
| | MLC | DLC | All LC | MLC | DLC | All LC |
| Rule-based LC model | 1.21 | 3.28 | 1.90 | 1.58 | 2.50 | 1.95 |
| Utility-based LC model | 1.07 | 4.71 | 2.28 | 1.33 | 3.13 | 2.05 |
| Data-driven LC model (no physical conditions) | 0.92 | 4.14 | 2.00 | 0.92 | 2.63 | 1.60 |
| LC GAN | 0.85 | 2.14 | 1.28 | 0.58 | 1.25 | 0.85 |

Note: each block is approximately 6 meters.

The experimental results indicate that estimation errors for MLCs are consistently lower than those for DLCs across both datasets. This can be attributed to the stronger behavioral consistency and more concentrated spatial distribution of MLCs, which result in better estimation performance across all models. The rule-based and utility-based LC models, which possess high interpretability, perform well in estimating MLCs, as MLCs tend to occur in relatively predictable spatial zones under less complex traffic conditions. However, their performance deteriorates for DLCs, which exhibit higher spatial dispersion and behavioral variability. Unlike LC behaviors on freeways, LC behaviors in arterial environments are strongly influenced by traffic signals and intersection geometry. Although rule- and utility-based models can provide certain filtering mechanisms to eliminate implausible lane-change positions, they fail to fully account for these context-dependent effects, resulting in reduced accuracy in complex arterial settings.

While data-driven LC models are theoretically capable of identifying behavioral patterns in LC by learning from data distributions, the tested data-driven LC model without physical conditions also struggle to capture the high randomness of DLCs. For instance, in the NGSIM dataset, the data-driven model even underperforms the rule-based model in estimating DLCs.

In contrast, the proposed LC-GAN effectively integrates the complementary strengths of the aforementioned approaches, achieving superior estimation performance across both datasets and for both types of LCs. By leveraging a conditional generative adversarial architecture, the LC-GAN combines the capability of deep learning models to capture complex behavioral dependencies with the interpretability of physical conditions. Specifically, the model incorporates safety-related physical conditions, such as acceptable gaps to surrounding vehicles ($C_{safety}$) and efficiency-related physical condition, such as preference for lane with shorter queue ($C_{signal}$) as physical conditions, guiding the generative process toward physically feasible and behaviorally realistic outcomes. These conditions not only enhance behavioral realism but also stabilize adversarial training and improve learning efficiency, allowing the model to compensate for the limitations of small-sample trajectory. Consequently, the proposed LC-GAN achieves an average estimation error of less than 1.3 blocks (approximately 8 meters) for all LC types across both real-world datasets, demonstrating its promising potential as a robust framework for LC behavior modeling in arterial environments.

### 4.2.2 Ablation experiment for LC GAN

To further validate the impact of the physical conditions embedded in the LC-GAN, we conducted an ablation experiment by sequentially removing each physical condition $C_{safety}$, $C_{signal}$ and $C_{geo}$. The results are summarized in the bar chart in Fig. 10, where the black numeric labels indicate the block error for each ablated model, and the white percentages represent the corresponding performance degradation relative to the proposed LC-GAN.

The results reveal that the removal of any single condition component leads to a noticeable degradation of performance. Specifically, $C_{geo}$ which classifies different LC types (e.g., DLC and MLC), plays a pivotal role in estimation accuracy. Its absence leads to a significant decline in estimation performance. $C_{safety}$, which enforces strong conditions against unsafe LC in critical blocks, has the most pronounced impact. Omission of this condition resulted in a substantial increase in both DLC and MLC estimation errors. $C_{signal}$, which prevents LC in area with queued or stationary vehicles, causes a moderate increase in error when omitted.

Overall, this ablation study underscores that the integration of physical conditions, rather than the adversarial architecture alone, is the key driver in improving training efficiency in the proposed framework. Each physical condition, safety ($C_{safety}$), signal-control ($C_{signal}$), and geometric condition ($C_{geo}$), contributes uniquely to the performance of the LC GAN, and that the joint integration of all three elements yields the most accurate and robust LC trajectory reconstruction.

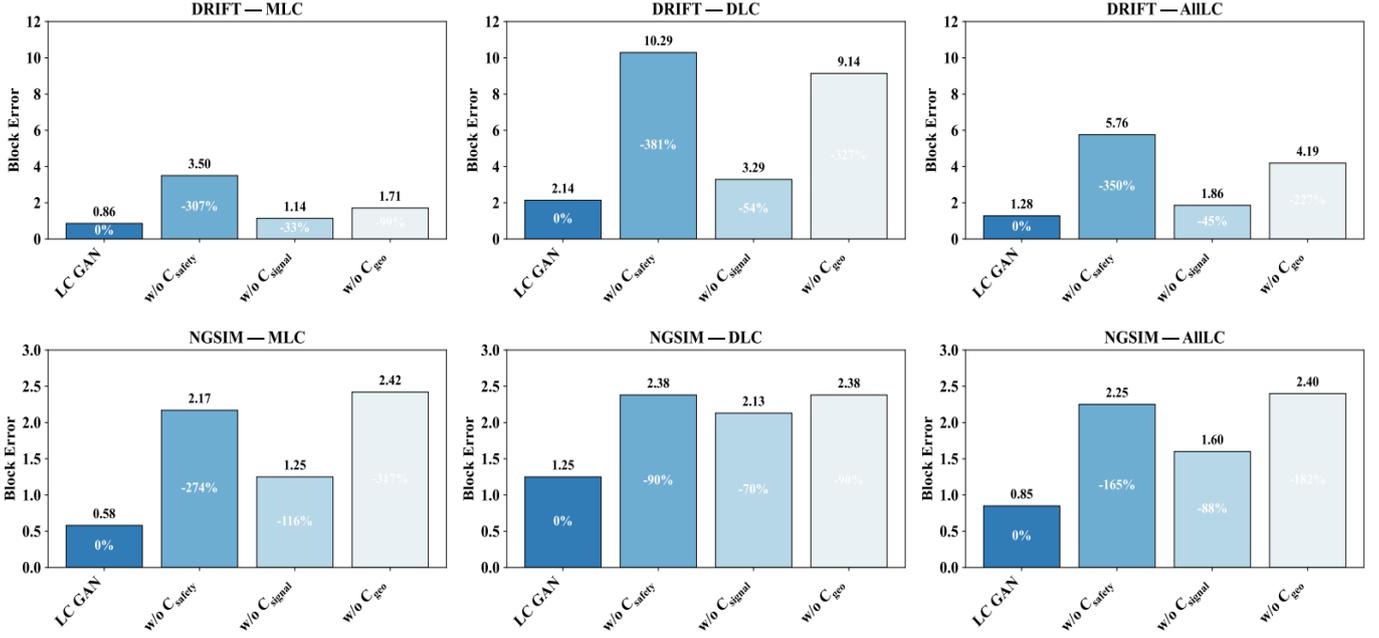

Fig. 10 Results of removing different condition information

## 4.3. Complete Traffic Flow Time-Space Diagram

Fig. 11 and Fig. 12 present a comparative analysis of trajectories reconstructed by the proposed framework against the corresponding ground truth across multiple lanes, using test-cycle data from the DRIFT and NGSIM datasets, respectively. In these figures, reconstructed LC and non-LC trajectories are depicted as yellow and green dashed lines, respectively, while the ground truth trajectories are represented by grey dashed lines. LCPs are explicitly marked with red stars.

In the DRIFT dataset, LCPs are predominantly concentrated near the ramp area, located approximately between 750 m and 1200 m from the origin, influenced by the connected ramp. Most of these trajectories correspond to secondary LC behaviors from the ramp-connected lane (Th1) to the left-turn lane (LT), representing typical MLCs. Vehicles entering via the ramp rapidly identify safe gaps and execute LCs, resulting in closely spaced LCPs within individual trajectories, thereby enhancing LCP estimation accuracy for MLC vehicles. In the adjacent through lanes (Th1 and Th2), DLC vehicles, such as veh1 and veh2, transition from the more congested Th2 to the less queued Th1. As shown for veh1, the LC decisions are typically made near the queue tail, where drivers can sufficiently assess the queuing conditions in both current and target lanes. The LCP is estimated accurately under the condition of $C_{signal}$ in our proposed framework. Meanwhile, veh2, which encounters no leading vehicle upon entering the lane, promptly evaluates traffic conditions in both lanes and executes an early LC near the origin. This behavior is accurately captured by our framework by introducing the dual physical conditions of $C_{safety}$ with $C_{signal}$. Additionally, veh3 highlights a scenario where a DLC vehicle makes a LC early from Th1 to Th2. As veh3 follows a slower leading vehicle and encounters insufficient safe spacing to maintain its original speed in Th1, it changes lanes into Th2, where the preceding vehicle moves faster, thus allowing veh3 to maintain higher speed and a safer headway. This adaptive behavior is effectively reconstructed by our framework, guided by the $C_{safety}$. Overall, these case-specific analyses highlight the ability of the proposed method to accurately reproduce realistic LC behaviors under varying physical conditions.

In the NGSIM dataset, the distribution of LC positions is more dispersed but still exhibits a trend analogous to those observed in the DRIFT dataset. Specifically, MLCs tend to concentrate near the upstream boundary of the study segment, while DLCs show greater stochasticity. As illustrated by veh4, veh5, and veh6, vehicles frequently perform LCs near the queue tail in an attempt to join shorter queues. However, veh7 transitions from Lane 2 (with a shorter queue) to Lane 3 (with a longer queue), making it difficult to infer the underlying motivation for such LC behavior. Since these vehicles are not ced by gaps to preceding vehicle in the target lane, the estimated LCP are closer to the segment's entry point rather than the queue tail, resulting in certain estimation inaccuracies. Nevertheless, owing to their limited disruptive impact on surrounding traffic, the resulting LCP errors remain within an acceptable range across the arterial network.

As illustrated in the Fig. 11 and Fig. 12, the empirical trajectories (gray dashed lines) from both real-world datasets exhibit car-following behaviors with natural stochasticity and dynamic variations. Minor discrepancies between reconstructed and observed trajectories are mainly attributable to the inherent variability in real-world free-flow driving behaviors. Despite these nuances, the proposed framework successfully reconstructs heterogeneous driving patterns across different vehicle platoons. Supported by the LC GAN, the framework achieves comprehensive and coherent LC trajectories reconstruction.

Furthermore, this study introduces a systematical framework that integrates the LC-GAN and Trajectory-GAN, effectively incorporating LC dynamics into trajectory reconstruction process. During training, the LC-GAN first estimates LC probabilities across a feasible range of positions, capturing the inherent uncertainty in LC position estimation. Subsequently, since the output of the LC GAN directly influences the conditional input of the Trajectory-GAN, the reconstruction loss from the Trajectory-GAN is back propagated to refine the LC position estimates. The Trajectory-GAN then narrows the feasible range of LC positions by incorporating kinematic consistency and safety condition. Through iterative joint training between the two models, this framework not only improves the accuracy of LC estimation but also enhances the overall accuracy of trajectory reconstruction. For instance, in the case of Veh6 from the NGSIM dataset, Trajectory-GAN infers the stopping position and adjusts the LC position accordingly. An early LC would introduce a collision risk with the inferred stopping point, whereas an LC closer to the queue ensures both safety and smooth trajectory integration, thereby reducing estimation error.

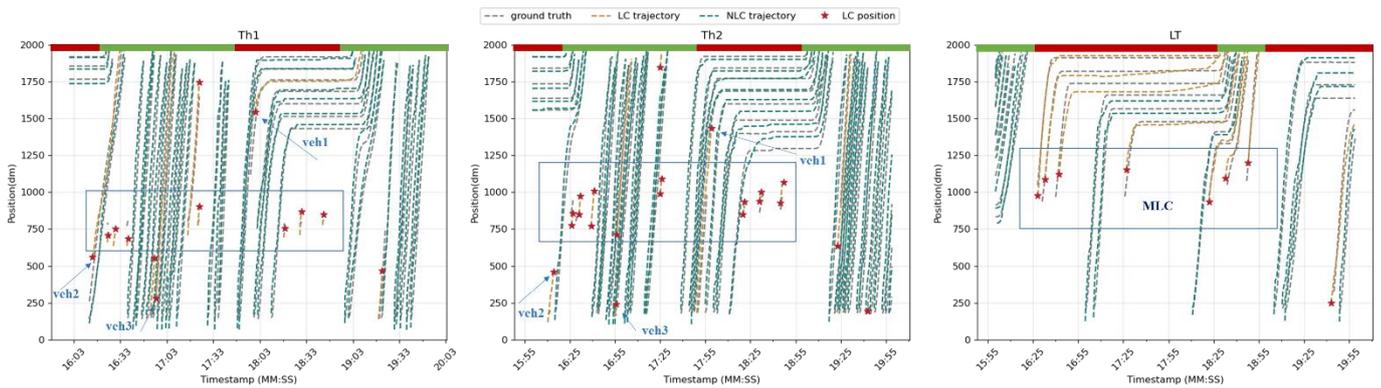

Fig. 11 The comparison between reconstructed trajectories (yellow dotted line represent LCVs, and green dotted line represent non-LCVs) and the real collected trajectories (grey dotted line) under different lanes in DRIFT dataset

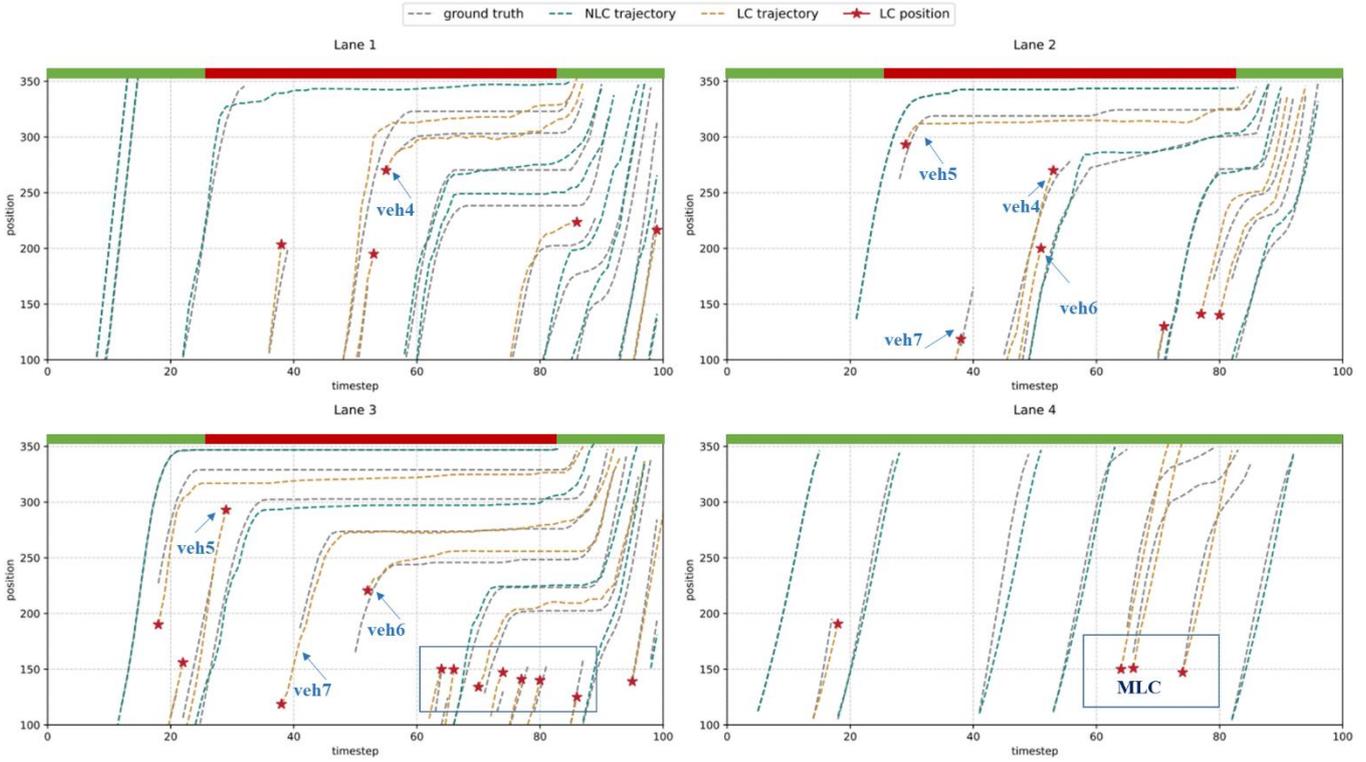

Fig. 12 The comparison between reconstructed trajectories (yellow dotted line represent LCVs, and green dotted line represent non-LCVs) and the real collected trajectories (grey dotted line) under different lanes in NGSIM dataset

## 4.4. The benefits of the integrated framework of LC-GAN and Trajectory-GAN

A systematic comparison was conducted between the integrated (joint training) framework and the sequential (separate training) framework for the LC-GAN and the Trajectory-GAN. The joint training strategy optimizes both networks simultaneously within a unified learning process, while the separate training approach independently optimizes each model. The comparative results are presented in Fig. 13, which consists of eight subplots illustrating the evaluation outcomes of the two models across both datasets, based on four metrics: Block Error (BE), Queue Error (QE), Time Error (TE), and Position Error (PE). Dashed lines indicate the improvement ratio achieved by the integrated framework over the sequential one.

QE, which measures the discrepancy in stopping positions between the reconstructed and ground-truth trajectories, shows minimal sensitivity to the training strategy, as LCs rarely occur precisely within the queue area. However, the integrated framework effectively reduces the feasible range of LCPs, resulting in a noticeable improvement (26%) in BE. This improvement is attributed to the integrated framework's ability to significantly enhance the accuracy of LCP estimation, enabling precise segmentation of hypothetical trajectories in both the current and target lanes. Consequently, this refinement improves the accuracy of the initial trajectory estimation for LC vehicles, facilitating more effective refinement during the Trajectory-GAN phase, and leading to overall improvements in both TE (around 35%) and PE (around 16%).

The enhanced performance can be attributed to the synergistic learning effects facilitated by joint optimization, which fosters coherent feature representation between LC behavior modeling and full-trajectory reconstruction. In contrast, the sequential framework fail to fully capture the intricate interdependencies between LC behavior and continuous trajectory dynamics, resulting in suboptimal reconstruction fidelity and behavioral inconsistencies. These findings underscore the importance of integrated, joint learning architectures for multi-task traffic trajectory reconstruction and validate the efficacy of the proposed integrated framework.

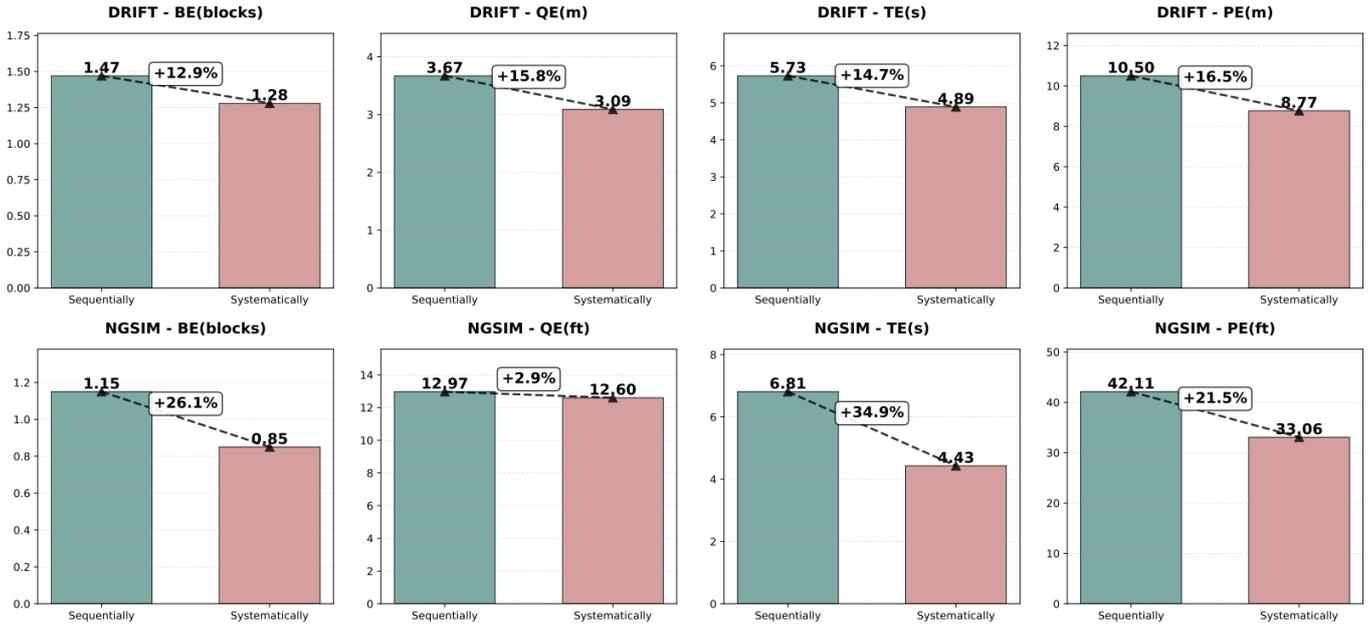

Fig. 13 Comparative evaluation of systematical versus sequential framework frameworks for LC GAN and Trajectory GAN

## 4.5. Performance comparison with other baselines

To evaluate the rationale for integrating the LC-GAN with the Trajectory-GAN in the proposed framework, this study establishes four comparative baseline models that sequentially connect the LC model and the car-following model. Additionally, a state-of-the-art data-driven trajectory prediction method with a joint connection structure is included as a fifth baseline. The detailed configurations of all models are summarized in Table 2.

The reconstruction performance was assessed by comparing all reconstructed trajectories against the corresponding ground-truth trajectories within the testing cycle using multiple error metrics: Block Error (BE) for LCP estimation, and Queue Error (QE), Time Error (TE), and Position Error (PE) for overall trajectory reconstruction performance. To account for the inherent stochasticity of the task, the mean values (triangular markers) and standard deviations (vertical bars) of these metrics are reported in Fig. 14.

Table 2. The detailed configurations of proposed framework and the comparative baselines

| Model | Lane Change Algorithm | Trajectory reconstruction Algorithm | Connection |
|---|---|---|---|
| Model I | Data-driven LC model (DBN) | Physical car following model (IDM) | Sequentially |
| Model II | Data-driven LC model (DBN) | Data-driven car following model (LSTM) | Sequentially |
| Model III | Data-driven LC model (DBN) | Trajectory-GAN | Sequentially |
| Model IV | LC-GAN | Physical car following model (IDM) | Sequentially |
| Model V | Data-driven trajectory prediction model without physical conditions (Meng et al., 2023) | | Jointly |
| Model VI | LC-GAN | Trajectory-GAN | Jointly |

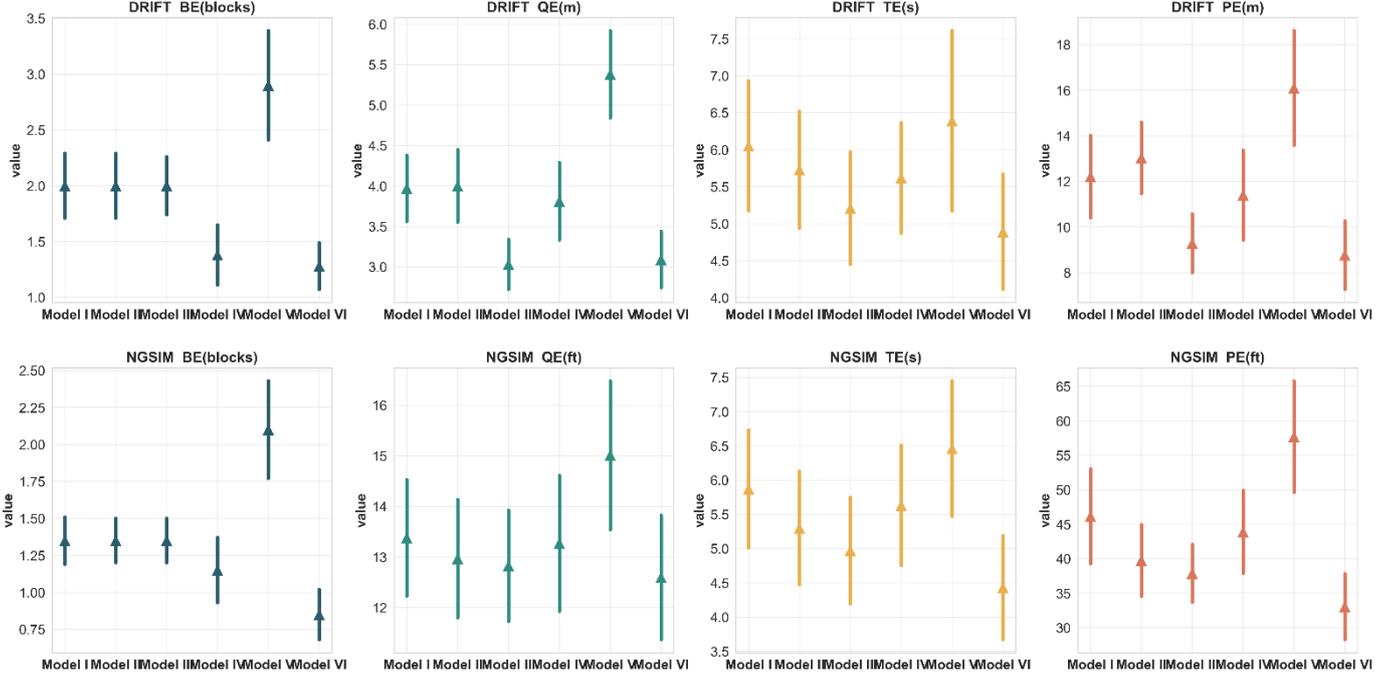

Fig.14 The reconstruction performance (BE, QE, TE and PE) of diverse baselines

The experimental results reveal the following key findings. In terms of overall performance, the integrated framework (Model VI) proposed in this study demonstrates the best performance, followed by Model III, Model II, Model IV and Model I, with Model V yielding the least favorable performance.

Under appropriate physical conditions, LC-GAN and Trajectory-GAN exhibits better performance compared to typical data-driven approaches without physical conditions and physical-based models. This is reflected in performance metrics: Model IV and VI with LC GAN achieve superior BE, while Model III and VI with trajectory-GAN demonstrate lower QE, TE and PE. By integrating the LC-GAN with Trajectory-GAN, the framework effectively captures the inherent stochasticity of LC behavior and dynamic driving characteristic, generating complete trajectories that closely align with real-world observations. In contrast, data-driven LC models rely on large, high-quality data, and when samples are limited, models such as DBN fail to represent complex LC behaviors, leading to higher BE, TE and PE values, as observed in Model I and II compared with Model V. Model III, which also uses a DBN for LC modeling, exhibits similar high BE to Model I and II yet achieves lower TE and QE through integration with the Trajectory-GAN. Model II, utilizing a data-driven car-following model (LSTM) without GAN architecture or physical conditions, demonstrates no distinct advantage over the Model I with a physics-based car-following model. These results confirms that physical conditions provide valuable guidance for GAN-based approaches, enhancing both performance and reliability.

While the physics-based models ensures that vehicle movements comply with kinematic constraints, it fails to fully capture the dynamic variations across vehicles under different signal phases or account for individualized driving characteristics. Although its combination with LC GAN (Model IV) yields relatively satisfactory BE performance, both PE and TE remain substantial. When sufficient historical data are available, data-driven trajectory prediction methods (e.g., Model V in this paper) may achieve desirable performance in short-term trajectory prediction (Meng et al., 2023). However, under complex LC and dynamic driving conditions, such methods does not explicitly distinguish between LC behavior modeling and trajectory reconstruction. Moreover, it struggles to account for inter-vehicle coordination and game-theoretic interactions, particularly in small-sample settings, resulting in the poorest estimation performance among all models.

Analysis of standard deviations across different models reveals that physics-based models yield more concentrated prediction distributions. In contrast, data-driven and GAN-based approaches exhibit greater variability, with purely data-driven methods, which directly model LC behavior and trajectory reconstruction together showing the highest dispersion. The proposed framework (Model VI), which effectively integrates physical principles with generative adversarial training, achieves not only the lowest mean error but also maintains a relatively narrow and acceptable deviation range. This balance between accuracy and stability establishes a solid foundation for practical applications combining LC behavior modeling and trajectory reconstruction.

## 5. Conclusion

Urban arterial networks, accommodating both human-driven vehicles (HVs) and connected vehicles (CVs), present opportunities to leverage sparse CV data for traffic state estimation and operational optimization. In this context, this study develops a Multi-task joint Generative Learning-based Trajectory Reconstruction Framework (MGL-TRF) that integrates an LC-GAN and a Trajectory-GAN to reconstruct complete time-space diagrams across multiple lanes at arterial intersections.

By incorporating physical conditions, including safety, signal control and geometric configuration, the LC-GAN accurately infers dynamic lang-changing (LC) behavior and remains reliable even under limited-sample conditions. Meanwhile, the Trajectory-GAN, enhanced by a physics-based car-following model, generates behaviorally plausible trajectories that adhere vehicle dynamics while adapting to varying traffic conditions. A key contribution of this study lies in the multi-task joint generative learning, which jointly models car-following and LC behaviors by leveraging their interactions as mutual auxiliary supervision and physical conditions. This design ensures both the physical plausibility and systemic integrity of the reconstructed trajectories.

Validation on two real-world datasets (DRIFT and NGSIM) demonstrated that the proposed framework outperforms the state-of-the-art baselines, including rule-based, utility-based, and purely data-driven models, achieving high accuracy for both mandatory and discretionary LC scenarios. Comparative and ablation experiment further confirm the multi-task joint learning and physics-informed conditions significantly enhance model performance. Overall, the MGL-TRF offers an effective and scalable solution for high-fidelity trajectory reconstruction in mixed-autonomy traffic environments.

For future research, several directions are suggested. First, for traffic cycles where no CV trajectory is available in a lane, the framework could be extended by incorporating additional fixed detector information and enable more comprehensive network-wide traffic trajectory reconstruction. Such a hybrid approach holds particular promise for urban networks with low CV penetration rates. Second, for scenarios where vehicle arrival and departure times are unavailable, which prevents direct identification of LC vehicles, future work may integrate higher-level CV sensing or cooperative perception data to infer LC behavior more effectively. Third, applying transfer learning and online adaptation techniques could allow the proposed framework to generalize across intersections with different geometric and control characteristics, enhancing its scalability for citywide deployment.

## CRediT authorship contribution statement

**Mengyun Xu**: Conceptualization, Methodology, Data curation, Software, Formal analysis, Investigation, Writing – original draft, Writing – review & editing. **Jie Fang**: Conceptualization, Methodology, Data curation, Software, Writing – original draft, Writing – review & editing. **Eui-Jin Kim**: Conceptualization, Methodology, Project administration, Supervision, Writing – review & editing. **Tony Z. Qiu**: Conceptualization, Project administration, Writing – review & editing. **Prateek Bansal**: Conceptualization, Validation, Investigation, Writing – review & editing.

## Declaration of competing interest

The authors declare that they have no known competing financial interests or personal relationships that could have appeared to influence the work reported in this paper.

## Declaration of generative AI and AI-assisted technologies in the writing process

During the preparation of this work, the author(s) used ChatGPT-4o (OpenAI) to improve the readability and clarity of the manuscript. After using this tool, the author(s) reviewed and edited the content as needed and take full responsibility for the content of the published article.

## Data availability

The data supporting this study will be made publicly available in an online repository upon the acceptance of the manuscript.

## Acknowledgements

This research was conducted at the Department of Civil and Environmental Engineering at National University of Singapore (NUS) when the first author was an exchange student, supported by the Chinese Scholarship Council (CSC) scholarship. Prateek Bansal was supported by the Presidential Young Professorship. Authors also acknowledge support from the National Natural Science Foundation of China under Grants 52172332 and 71901070. This work was supported by the National Research Foundation of Korea (NRF) grant funded by the Korea government (MSIT) (No.RS-2024-00337956).